\documentclass[aps,epsfig]{revtex4}
\usepackage[varg]{txfonts} 
\usepackage[latin1]{inputenc}
\begin{document}

\title{Exact Analytical Solutions in Three-Body Problems and Model of Neutrino Generator}

\author{Nurgali Takibayev}

\address{Department of Physics \& Mathematics, \\
Kazakh National Pedagogical University, Almaty, Kazakhstan}

\email{teta@nursat.kz}

\begin{abstract}

Exact analytic solutions are obtained in three-body problem for the scattering of light particle on the subsystem of two fixed centers in the case when pair potentials have a separable form.
Solutions show an appearance of new resonance states and dependence of resonance energy and width on distance between two fixed centers. The approach of exact analytical solutions is expanded to the cases when two-body scattering amplitudes have the Breit-Wigner's form and employed for description of neutron resonance scattering on subsystem of two heavy nuclei fixed in nodes of crystalline lattice. 
It is shown that some resonance states have widths close to zero at certain values of distance between two heavy scatterer centers. It gives the possibility of transitions between states. One of these transitions between three-body resonance states could be connected with process of electron capture by proton with formation of neutron and emission of neutrino.   
This exoenergic process leading to the cooling of star without nuclear reactions is discussed. 
\end{abstract}

\maketitle

\section*{Introduction}

Exact solutions are of importance in quantum mechanics since they give rise to plain comprehension into phenomena occurring in the systems being considered. Usually exact solutions in analytic form can be obtained in simple models and grown effective tools for investigation of more complicated problems. It should be noted that revealed features of model problems take place actually in appropriate real physical systems. 

Some of model problems are of the significant importance in quantum physics.  
So, there are well known the Thomas effect of collapse into a center in system of three identical particles whose pair interactions have a zero range \cite{TakibayevNZh_Thomas} and the Efimov effect of the condensation of levels in three particle spectrum and growth of their number as the ratio of the pair scattering length to the range of pair forces increases \cite{TakibayevNZh_Efimov}. Note the investigation of threshold anomalies in cross sections and phenomena of long-range characters in \cite{TakibayevNZh_BAZ,TakibayevNZh_TakPen94}.  

In nuclear physics, for example, solutions to specific physical problems are usually obtained by means of cumbersome numerical calculations. 
At the same time, the presence of model problems that admit solutions in an analytic form could provide the possibility of performing an exact analysis of many phenomena inherent in system of complicated internal structure. A determination and investigation of such model problems is the main objective of the present study.

In model problem of light particle scattering on two heavy fixed centers analytic solutions demonstrate how the new resonances appear in three-body system. They show that positions and widths of these resonances depend on distance between two centers. 

Analytic solutions for the problem of light particle scattering on two heavy particles are found if two associated simplifications are acting in the system \cite{TakibayevNZh_PAN08}: 
\newline
- the limit of $\zeta = m/M \rightarrow 0$, where $m$ - mass of light particle and $M$ - mass of heavy particles, identical for simplicity; 
\newline 
- and pair $t_{i}$-matrices have a separable form ($i$ - number of pare): 
$t_{i} = |\nu_i > \eta_i < \nu_i |$. 

This model is used for description of resonance scattering in system where the light particle is a neutron and two nuclei are placed in fixed centers. 

In the case when two-body neutron-nucleous scattering amplitudes have the resonance Breit-Wigner's form analytic solutions are also obtained and  
three-body scattering amplitudes manifest the resonance behavior, too. 

Moreover, solutions for amplitudes of neutron scattering on subsystem of two fixed heavy nuclei show that sets of new resonance states appear in the three-body system. It is remarkable that positions and widths of resonances depend on $b$ - distance between two heavy scatterer centers, and there are values of $b$ when resonance widths come to zero. It seems that these quasi-bound states must have very big lifetimes. 

The model of subsystem consisted of two $\alpha$-particles taken as scatterer centers is considered to investigate neutron and proton scattering amplitudes. The comparison of their resonance states displays that some proton states are situated on energy scale  more than $0.8$ MeV above the neutron ones. 

In this case the reactions of electron capture by proton with emission of neutrino become possible. It gives cause for consideration  of neutrino generator model. 

The proposal of experiment is suggested to investigate  behavior, appearance and parameters of new three-body resonances, for instance, with using of monocrystals under high pressure and low temperature.  
Some applications of the exact solution method interesting in physics are discussed.

\section{Three-body problem with two-body separable interactions}
\label{TakibayevNZH_sec:2}
A mathematically rigorous solution to three-body problems was given by Faddeev \cite{TakibayevNZh_Faddeev}. The set of Faddeev equations for the $T$-matrix elements can be written in the form:
\begin{equation}
  \label{TakibayevNZh_eq:1}
  T_{i,j} = t_i \: \delta_{ij} + \sum t_i \: G_0 (Z) \: \bar{\delta_{i,l}} \: T_{l,j} 
\end{equation}
where  $t_i = V_i + V_i G_0 (Z) \: t_i  $  are pair $t$ matrices associated with pair interaction potentials  $V_i $  and determined in the space of three particles, ($i,j,l = 1,2,3$), $G_0 $ - the Green's function for three free particles, $Z$  is  a parameter that characterizes the problem and which corresponds to $E$ - the total energy of the three-particle system, and $\bar{\delta_{i,j}} = 1 - \delta_{i,j} $. 
The total $T$-matrix is the sum over the indices $i$ and $j$, $T = \sum T_{i,j}$. 

Let us consider the example of simple pair separable potentials $V_{i} = |\nu_{i}> \lambda_i < \nu_{i} |$ , where $ \lambda_i $  is a pair coupling constant. Sandwiched between free wave functions, the operator $V_i $  assumes the conventional form of a function of coordinates or momenta, depending on the choice of representation for these wave functions. For such potentials, pair $ t_i $- matrices are completely determined in an analytic form \cite{TakibayevNZh_Newton} 
\begin{equation}
  \label{TakibayevNZh_eq:2}
  t_i = |\nu_i> \eta_i < \nu_i | ,  
\end{equation}
where
\begin{equation}
  \label{TakibayevNZh_eq:3}
  \eta_i^{-1} = \lambda_i^{-1} - < \nu_i | G_0^+(E) |\nu_i >  . 
\end{equation}
Sometimes, they are referred to $\eta_i $ as enhancement factors; in this way, the similarity of the shapes of the potential and amplitude and the dependence of the amplitude on $ \lambda_i $   and the energy of the subsystem are emphasized.

In the common case short-range pair interactions can be represented as sums of separable members   
\begin{equation}
  \label{TakibayevNZh_eq:4}
  V_i = \sum_n |\overleftarrow{\nu}_{i,n}> \lambda_{i,n} < \overrightarrow{\nu}_{i,n} | ,  
\end{equation}
where $n$ can be a set of indices and contains the number of separable member and also discrete quantum numbers $L,S,J$.  

In~(\ref{TakibayevNZh_eq:4}) overarrows point out the presence of spherical function, for example
\begin{equation}
  \label{TakibayevNZh_eq:5}
 < \overrightarrow{\nu}_{i,n}|\vec{q}> = \nu_{i,n}(\vec{q}) =\nu_{i,n}(q)| \cdot Y^M_L (\hat{\vec{q}}) . 
\end{equation}
Therefore $t$-matrix can be written in the form 
\begin{equation}
  \label{TakibayevNZh_eq:6}
  t_i = \sum_{n,m} |\overleftarrow{\nu}_{i,n}> \eta_{i;n,m} < \overrightarrow{\nu}_{i,m} | ,  
\end{equation}
where
\begin{equation}
  \label{TakibayevNZh_eq:7}
  \eta_{i;n,m}^{-1} =  \delta_{nm} \lambda_{i,n}^{-1} - < \overrightarrow{\nu}_{i,n} | G_0 (E) |\overleftarrow{\nu}_{i,m}> .  
\end{equation}

For simplicity, we come back to simple form~(\ref{TakibayevNZh_eq:2}), supposing that transition to the complicated form~(\ref{TakibayevNZh_eq:6}) does not meet  any difficulties.   

Introducing $P_{ij}$ matrix by form 
\begin{equation}
  \label{TakibayevNZh_eq:8}
 T_{i,j} = t_i \: \delta_{ij} +  |\nu_i >  \eta_i \; P_{i,j} \; \eta_j < \nu_j | \ ,  
\end{equation}
we arrive at  
\begin{equation}
  \label{TakibayevNZh_eq:9}
  P_{i,j} = \Lambda_{i,j} + \sum \Lambda_{i,k} \; \eta_{k} \; P_{k,j} \ , 
\end{equation}
where $ \Lambda_{i,j} = < \nu_i | G_0^+(E) |\nu_j > $, $ i \neq j$ .

It is important that $ \Lambda_{i,i} \equiv 0 $ -  i.e. the diagonal elements vanish identically. This distinctive feature of Faddeev equations ensures the compactness of kernels in respective integral equations and the existence and uniqueness of solutions. Indeed, the singularities of the Born terms become weaker upon successively iterating Faddeev equations, while the kernels of the integral equations become normalizable; therefore, solutions exist and are obtainable \cite{TakibayevNZh_Faddeev,TakibayevNZh_FadMerk}.

In common case (see, for example \cite{TakibayevNZh_PAN08}) when the two-body interaction between heavy particles exists we can determine "the nuclear equation"
\begin{equation}
  \label{TakibayevNZh_eq:10}
  P_{1,1'} = V_{1,1'} + \sum_{1"} V_{1,1"} \; \eta_{1"} \; P_{1",1'} . 
\end{equation}
Here the effective potential between heavy particles
\begin{equation}
  \label{TakibayevNZh_eq:11}
  V_{1,1'} = \sum_{k, l = 2,3} \Lambda_{1,k} \; \eta_{k} \; (\delta_{kl} +  M_{k,l}) \; \eta_{l} \; \Lambda_{l,1'} , 
\end{equation}
can be determined by means of "electronic equation":
\begin{equation}
  \label{TakibayevNZh_eq:12}
  M_{k,l} = \Lambda_{k,l} + \sum_{\rho = 2,3} \Lambda_{k,\rho} \; \eta_{\rho} \;   M_{\rho,l}  . 
\end{equation}

Above-mentioned terms are given from the well-known Born-Oppenheimer approximation. This approximation is the assumption that the electronic motion and the nuclear motion in molecules can be separated. The electronic wave-function depends upon the nuclear positions $\vec{R}_{2,3}$ but not upon their velocities, i.e., the nuclear motion is so much slower than electron motion that they can be considered to be fixed.

In the case of fixed heavy centers we have to solve only Eq.~(\ref{TakibayevNZh_eq:12}), which gives total description of light particle scattering amplitude on these centers \cite{TakibayevNZh_PAN08}. 

\section{The scattering problem on two heavy centers}
\label{TakibayevNZh_sec:3}
Let us now consider the problem where one of the particles is light, while the other two are heavy. Specifically, we examine the limiting case of $\zeta = m/M \rightarrow 0$. 

The total energy of the system is $E = \sum p^{2}_{0i} /2m_i = p^2_0/2m $ , where $\vec{p}_{01} = \vec{p}_0  $ - being initial momentum of the light particle. The heavy particles are labeled with the indices of 2 and 3. For the sake of simplicity, we assume them to be identical.

In the limit  $\zeta \rightarrow 0$, the form factors for the pair potentials of interaction  between the light particle and any of the heavy particles ceases to depend on the heavy-particle momentum: $\nu(\vec{q}_{12} \rightarrow  \nu (\vec{p})$,  $\nu (\vec{q}_{13} \rightarrow  \nu (\vec{p}) $, since $ \vec{q}_{12} = (m_2 \vec{p}_1 - m_1 \vec{p}_2)/(m_2 + m_1) \rightarrow \vec{p}_1 = \vec{p} $ and, accordingly, $ \vec{q}_{13} = (m_3 \vec{p}_1 - m_1 \vec{p}_3)/(m_3 + m_1) \rightarrow \vec{p}_1 = \vec{p} $.

The enhancement factors in the $t$ matrices for these pairs become functions only from the initial energy of the light particle - that is, they are functions of its initial momentum: $ \eta_2 = \eta_3 \rightarrow \eta(p_0)$.

In the limit $\zeta \rightarrow 0 $ the expression for $ \Lambda_{k,l} $ in "electronic equation"~(\ref{TakibayevNZh_eq:12}) is   
\begin{equation}
  \label{TakibayevNZh_eq:13}
  \Lambda_{2,3} = 2m \frac{\nu_2 (\vec{p}) \nu_3 (\vec{p})}{p^2_0 - p^2 + i0} \equiv f(p_0,\vec{p}) . 
\end{equation} 
Taking into account the conservation of total momentum in three-particle system $\vec{p} = - \vec{p}_2 - \vec{p}'_3 $ , we represent the potential  $ \Lambda_{2,3} $  in the integral form
\begin{equation}
  \label{TakibayevNZh_eq:14}
  \Lambda_{2,3}(\vec{p}_2, \vec{p}'_3) = \int d\vec{r} \exp(i \vec{r}\vec{p}_2) J_{2,3}(p_0; \vec{r}) \exp(i \vec{r}\vec{p}'_3) , 
\end{equation} 
where
\begin{equation}
  \label{TakibayevNZh_eq:15}
  J_{2,3}(p_0; \vec{r}) = \int d\vec{p} \exp(i \vec{r}\vec{p}) f(p_0; \vec{p})  . 
\end{equation} 

In Eq.~(\ref{TakibayevNZh_eq:14}), we label heavy-particle variables at the exit from interaction region with a prime; at the entrance, they carry no primes. We will use this notation below.

Then we determine the Fourier transform of the solution $ M_{k,l} \equiv M_{k,l}(p_0; \vec{p}_k \vec{p}'_l) \rightarrow   M_{k,l}( \vec{r}, \vec{r}') $, namely
\begin{equation}
  \label{TakibayevNZh_eq:16}
 M_{k,l} ( \vec{r}, \vec{r}')   
 = \int d\vec{p}_k d\vec{p}'_l 
   M_{k,l}\exp(i \vec{r}' \vec{p}'_l - i \vec{r}\vec{p}_k) . 
\end{equation} 
Take into consideration the relation~(\ref{TakibayevNZh_eq:14}) we obtain from Eq.~(\ref{TakibayevNZh_eq:12}) equation  
\begin{eqnarray}
  \label{TakibayevNZh_eq:17}
  M_{k,l} ( \vec{r}, \vec{r}') =  J_{k,l}(p_0; \vec{r}) \delta (\vec{r} + \vec{r}') +   \nonumber\\
+ \sum_\rho J_{k,\rho}(p_0; \vec{r})M_{\rho, l} (- \vec{r}, \vec{r}') . 
\end{eqnarray} 
Since the delta function removes the integration on the right-hand side of~(\ref{TakibayevNZh_eq:17}), the equation for $M_{k,l} ( \vec{r}, \vec{r}')$  is reduced to the extremely simple form.  

As a result, the solution of light particle scattering problem on two heavy centers  can be represented in analytical form 
\begin{equation}
  \label{TakibayevNZh_eq:18}
  M_{i,j} ( \vec{r}, \vec{r}') =  \frac{1}{I - B_{i,i}(\vec{r})}K_{i,j}(\vec{r}, \vec{r}') , 
\end{equation} 
where $i,j = 2,3$  - numbers of heavy centers, $ \vec{r}$ - the radius-vector according to the position of initial scattering center and $ \vec{r}' $ - the radius-vector according to the position of final scattering center  in c.m. of system.

Elements of diagonal matrix $B_{i,i}$ ($B_{i,j} = 0$ if $j\neq i$) are 
\begin{equation}
  \label{TakibayevNZh_eq:19}
  B_{i,i} ( \vec{r}) = \sum_{k=2,3} J_{i,k}(p_0; \vec{r}) \eta_k (p_0) J_{k,i}(p_0; - \vec{r})  \eta_i (p_0) . 
\end{equation} 
Elements of matrix  $K_{i,j}(\vec{r}, \vec{r}') $ are given as 
\begin{equation}
  \label{TakibayevNZh_eq:20}
  K_{i,i} ( \vec{r}) = \sum_{k=2,3} J_{i,k}(p_0; \vec{r}) \eta_k (p_0) J_{k,i}(p_0; - \vec{r}) \: \delta (- \vec{r} + \vec{r}')  ,   
\end{equation} 
if $j=i$, and 
\begin{equation}
  \label{TakibayevNZh_eq:21}
  K_{i,j} ( \vec{r}) =  J_{i,j}(p_0; \vec{r}) \: \delta ( \vec{r} + \vec{r}'),   
\end{equation} 
where $j \neq i$ .

And we can write out  two modes of solution $  M_{i,j} ( \vec{r}, \vec{r}') $
\begin{equation}
  \label{TakibayevNZh_eq:22}
  M_{i,j} ( \vec{r}, \vec{r}') =   M_{i,j}^+ ( \vec{r})\: \delta ( \vec{r} + \vec{r}') + M_{i,j}^- ( \vec{r})\: \delta ( -\vec{r} + \vec{r}')   , 
\end{equation} 
where
\begin{equation}
  \label{TakibayevNZh_eq:23}
  M_{i,j}^+( \vec{r}, p_0) =  \frac{1}{I - B_{i,i}(\vec{r})}J_{i,j}(\vec{r},p_0) , 
\end{equation}  
and 
\begin{equation}
  \label{TakibayevNZh_eq:24} 
   M_{i,i}^- ( \vec{r}, p_0) =  \frac{1}{I - B_{i,i}(\vec{r})}B_{i,i}(\vec{r}, p_0) \eta^{-1}_i (p_0) , 
\end{equation} 
as   $M_{i,j}^-  = 0 $ if $j\neq i  $.

In the case when pair potentials are sums of separable terms, expressions~(\ref{TakibayevNZh_eq:13}) -~(\ref{TakibayevNZh_eq:24}) have to be considered as the matrix expressions with respect to additional indices.

We assume above that heavy particles are strictly fixed at points $\vec{R}_2$ and  $\vec{R}_3$ = $\vec{R}_2 + \vec{b}$. We introduce the wave functions for these centers of identical heavy particles in the form \cite{TakibayevNZh_PAN08}
\begin{equation}
  \label{TakibayevNZh_eq:25} 
   \Psi_{i}( \vec{r}, \vec{R}_i) = C \exp \left[- \frac{(\vec{r} - \vec{R}_i)^2}{2 \Delta^2}\right] , 
\end{equation} 
for heavy particle  is localized in a bounded region centered at the point   $\vec{R}_i$, $i=2,3$.

The obvious normalization condition $<\Psi_{i}|\Psi_{i}> = 1 $ gives $C^2 = (\Delta^2 \pi)^{-3/2}$. In order to determine the physical scattering amplitude, it is necessary to sandwich the expression for $T$ - matrix between the wave functions for initial and final states of the system $ <\Psi_{in}|T| \Psi_{f}> $ .

The structure of these functions is obvious, for example, $<\Psi_{in}| = <\chi_1 \Psi_{2} \Psi_{3}|$ , where  $ \chi_1 $  is the free wave function for the light particle. 

 It should be emphasized that positions of heavy particles are specified in the c.m. frame of all three particles. This concerns their coordinates and momenta. It is then obvious that $\vec{R}_2 = - \vec{b}/2 $ and  $\vec{R}_3 = \vec{b}/2 $.

The scattering amplitude of light particle on two heavy centers is determined by the form: 
\begin{equation}
  \label{TakibayevNZh_eq:26} 
 f (\vec{b}; p_0) = \sum_{i,j=2,3} <\Psi_{i}( \vec{r}, \vec{R}_i)|M_{ij} ( \vec{r},  \vec{r}') | \Psi_{j}( \vec{r}', \vec{R}_j)> .  
\end{equation} 

In the limit $\Delta \rightarrow 0$, this amplitude comes to the expression 
\begin{equation}
  \label{TakibayevNZh_eq:27} 
 f (\vec{b}; p_0) = M(\vec{b}/2, p_0) + M(- \vec{b}/2, p_0)   ,  
\end{equation} 
where $ M(\vec{b}/2) = M^+ (\vec{b}/2) + M^- (\vec{b}/2)$  as following from ~(\ref{TakibayevNZh_eq:23}) and~(\ref{TakibayevNZh_eq:24}).

\subsection{Zeros of $D$-function}
\label{TakibayevNZh_sec:31}

It is clear that zeros of $D$-function, where $D = \det(I - B)$, correspond to poles of three-body amplitude $M(\vec{b}/2; p_0)$ in complex plane of $p_0$  (see~(\ref{TakibayevNZh_eq:18})).  And we can determine paths of motion of these zeros in complex plane of energy. 

For example, we can take the simple pair potential of Yamaguchi form acting in $S$-wave which has form-factors
\begin{equation}
  \label{TakibayevNZh_eq:28} 
\nu(p) = Const/ (1 + p^2/\beta^2) \  ,  
\end{equation} 
where $Const = \sqrt{8\pi/(2 m \beta)}$, $\beta$ -  the inverse range of nuclear forces. 
In this case the enhancement factor is
 \begin{equation}
  \label{TakibayevNZh_eq:29} 
\eta^{-1}(p_0) = \lambda^{-1} + (1 - i k)^{-2}  \  ,  
\end{equation} 
 $ k = p_0 /\beta$. 
We take $\hbar =1, c = 1$, for simplicity. 

Then we can get 
\begin{equation}
  \label{TakibayevNZh_eq:30} 
 J = 2\frac{\exp(-\tilde{b}) - \exp(i\tilde{b} k)}{\tilde{b} (1 + k^2)^2} + \frac{\exp(-\tilde{b})}{(1 + k^2)}   ,  
\end{equation} 
where $ J = J_{ij} = J_{ji}$, $ \tilde{b} = b\beta/2 $. 

We note that, by definition, the coupling constant $\lambda $ is real-valued. This follows from the unitarity and microcausality conditions even at the two-particle level \cite{TakibayevNZh_KKT}. This leads to a relation between the coupling constant $\lambda $ and the coordinates of those points $k_{res} = k_R + i k_I$ in the complex plane of $k$
\begin{eqnarray}
  \label{TakibayevNZh_eq:31} 
 \frac{[1 + k_R^2 - k_I^2]^2 - 4 k_R^2 k_I^2} {\lambda} =  
 \left[1+k_R^2 - k_I^2\right] \exp(-\tilde{b}) + \nonumber\\
 + \frac{\exp(-\tilde{b}) 
   - \exp(- k_I \tilde{b})\cos(k_R \tilde{b})}{\tilde{b}/2} - \left[1-k_I^2\right]^2   \ \  , \ \    
\end{eqnarray} 
and
\begin{eqnarray}
  \label{TakibayevNZh_eq:32} 
   - \frac{2 k_{I}(1 + k_{R}^2 - k_{I}^2)}{\lambda} =
 1 - 
 k_{I}\left(1 + \exp(-\tilde{b})\right) +  \nonumber\\ 
 + \frac{ \exp\left(- k_{I} \tilde{b}\right) \sin \left(k_{R} \tilde{b}\right)}  {\left(k_{R}\tilde{b}\right)}  \quad .  \qquad 
\end{eqnarray}
Owing to this, the value of $k_I = 0$ is forbidden for any $k_R \neq 0 $ and real $\lambda \neq 0$, and of course positive $ \tilde{b}$. 

This means that the zeros of the function $D (\tilde{b}, k)$ cannot intersect the real axis of the complex plane of $k$, or, in other words, they cannot be in the physical region of scattering.  

The problem can be further simplified if the pair interaction between the light particle and each of the fixed centers is taken to be contact. 

The corresponding solutions follows from~(\ref{TakibayevNZh_eq:30}) upon going over to the limit $\beta \rightarrow \infty $ and fixed value of the quantity $\kappa_0$; $E_B = -\kappa_0^2/2m $. The function $D$ then takes the form
\begin{equation}
  \label{TakibayevNZh_eq:33} 
 D = 1 + 2\frac{\exp(ibp_0/2)}{b (\kappa_0 + i p_0)}   .  
\end{equation} 

Zeros of D-function give the values of $x_R$ and $x_I$, where $x = bp_0/2$, $x = x_R + i x_I $,   can be found from the algebraic equations \cite{TakibayevNZh_PAN08}: 
\begin{eqnarray}
  \label{TakibayevNZh_eq:34} 
 x_I = x_0 + \cos(x_R) \exp(-x_I) ,  \\
 x_R = -\sin(x_R) \exp(-x_I)   , \nonumber 
\end{eqnarray} 
where $x_0 = b\kappa_0/2 $.
In the case of $x_R = 0$  the relationship analogous~(\ref{TakibayevNZh_eq:34}) is well-known \cite{TakibayevNZh_BAZ}. 

In the last case, 
\begin{equation}
  \label{TakibayevNZh_eq:35} 
 x_I = x_0 + \exp(-x_I) , 
\end{equation} 
$x_I > x_0 $ - that is, the three particle system is always bound more strongly than the two-particle subsystem. This is the reason why the light particle can be bound by the system of two fixed centers even if it is underbound by one isolated center \cite{TakibayevNZh_BAZ}.

The existence condition for a bound three-particle state is $x_0 > - 1$ , which is possible at any $\lambda $ obeying the inequality $\lambda \leq 0 $ but at specific values of $b$ that satisfy the recursive relation~(\ref{TakibayevNZh_eq:35}).

As for quasistationary states, the values of $x_R$ and $x_I$ can be found for them from the set of equations~(\ref{TakibayevNZh_eq:34}). Here, the condition $x_I < 0$ always holds, so that the resonance poles always lie in the lower half-plane of the complex plane of $x$. 

 It is remarkable that energies and widths of three-body resonances depend on parameter $b$ - the distance between the scattering centers. This dependence is an important feature of the three-body system. It is clear that bound, virtual and quasi-stationary states will be moving in complex plane of $p_0 $   with changing of distance between scattering centers \cite{TakibayevNZh_PAN08}.

Note that in common case, for example, of more complicated pair potentials in $S$-wave or of higher partial waves, zeros of function $D$ can intersect the real complex plane of $k$. That is the some of resonance points can get $k_I = 0$ but $k_R \neq 0$ at specific values of $b$.  

\section{Corrections to exact analytical solutions}
\label{TakibayevNZh_sec:4}

Supposing that $M_{k,l}$ in~(\ref{TakibayevNZh_eq:7}) and~(\ref{TakibayevNZh_eq:12}) are already defined  (see above Sect.~(\ref{TakibayevNZh_sec:3})) we resort to estimation of corrections to exact analytical solutions.  

In order to determine these corrections one can use iterative method within the framework of Faddeev equations. Note that the convergence of iterative procedure has an exponential behavior, that is more fast in comparison with the  ordinary perturbation theory \cite{TakibayevNZh_FadMerk}.  

Also the method  of coupling constant evolution may be used to find corrections to energies of bound or resonant states  \cite{TakibayevNZh_KKT,TakibayevNZh_PAN05}. 

First of all we can find three-body wave functions. Following~(\ref{TakibayevNZh_eq:9}) and Lippmann-Schwinger equations we get wave functions in continuum  
 \begin{equation}
  \label{TakibayevNZh_eq:36} 
   |\Psi> = \sum_j \left(|\phi_i> \delta_{ij} + G_0 |\nu_i> \eta_i  P_{ij} \lambda_j <\nu_j| \phi_j > \right) \  ,  
\end{equation} 
where $| \phi_i >$  - two-body wave functions.

In the case of spectrum we can connect the  wave function of three-body state with the residue of $T$-matrix in the pole
 \begin{equation}
  \label{TakibayevNZh_eq:37} 
   T (E)_{E\rightarrow E_n} \rightarrow \frac{V|\Psi_n> <\Psi_n| V}{E - E_n}  \ .  
\end{equation} 
According to~(\ref{TakibayevNZh_eq:8}),~(\ref{TakibayevNZh_eq:18}) and~(\ref{TakibayevNZh_eq:19}), for instance, in $S$-wave this amounts to 
 \begin{equation}
  \label{TakibayevNZh_eq:38} 
 |\Psi_n> = G_0 \sum_i |\nu_i > R_n  \ , 
\end{equation} 
where $R_n = \sqrt{(E_n - E_{2,res})/(2J)}$ to the accuracy of phase. 

Here $E_{2,res}$ is the energy where  $\eta^{-1}_i = 0$. This energy corresponds to the two-body "resonance" state - it may be bound or virtual, or quasi-stationary one. 

Meanwhile $E_n = E_n (b/2; p_0)$ corresponds to the  resonance state of three-body system. $J = J_{ij}(E_n,b/2) = J_{ji}(E_n,b/2)$ as fixed heavy particles are identical.
 
Now we know solutions of our simple model, i.e. elements of $T$-matrix, wave functions and energies of states. After that we can construct scheme for determination of corrections. 

For example, in framework of coupling constant evolution method we can consider  the evolution of system with increasing of small parameter $\zeta$. Resonance energy shifts are determined by equation 
 \begin{equation}
  \label{TakibayevNZh_eq:39} 
 \frac{dE_n}{d\zeta} = \frac{<\Psi_n| H_{\zeta} |\Psi_n >}{<\Psi_n | \Psi_n >}  \ , 
\end{equation} 
where $ H_{\zeta} = dH/d\zeta $, and shifts of wave functions by  
 \begin{equation}
  \label{TakibayevNZh_eq:40} 
 \frac{d|\Psi >}{d\zeta} = \frac{1}{E - H}\left(H_{\zeta} - E_{\zeta}\right)|\Psi>  \ , 
\end{equation} 
$ E_{\zeta} = dE/d\zeta $ \cite{TakibayevNZh_KKT,TakibayevNZh_PAN05}. Exact analytical solutions are taken here as boundary conditions for $H$, $E_n$ and $ |\Psi> $  at limit $ \zeta\rightarrow 0 $.   

Moreover, we can take into account small motions of heavy centers in expression~(\ref{TakibayevNZh_eq:25}). Decomposing expression ~(\ref{TakibayevNZh_eq:26}) in respect of $\Delta \neq 0$ (but $\Delta/b << 1$) one can obtain that the amplitude  in resonance point  $\sim \Delta^{-1}$, and corrections to resonance energy and width are linear with $\Delta/b << 1$.

Putting $W_{k,i}= T_{k,i} - t_i \delta_{ki}$, we can write expressions   
\begin{eqnarray}
  \label{TakibayevNZh_eq:41} 
 W_{1,1} =  \sum_{k=2,3} t_1 G_0 (E) W_{k,1} , \nonumber\\
 W_{k,1} = |\nu_k > \eta_k \cdot \left( \rho_{k,1} + F_{k,1} \right)  ,  
\end{eqnarray} 
where $k\neq 1$, and
 \begin{eqnarray}
  \label{TakibayevNZh_eq:42} 
 \rho_{k,1} =  < \nu_k | \: G_0 (E) \cdot T_{1,1}  , \nonumber\\
 F_{k,1} = \sum_{l=2,3}M_{k,l} \cdot\eta_l \cdot \rho_{l,1}  .  
\end{eqnarray} 
Then we can take a pair interaction between two heavy centers in arbitrary form not only separable one. Writing  $W_{11} = V_1 \cdot\tilde{W}_{11}$, we obtain   
 \begin{equation}
  \label{TakibayevNZh_eq:43} 
 \tilde{W}_{11} =  V^{ef}_{11} + V^{ef}_{11} G^{-1}_1 G_0 \tilde{W}_{11}  \  ,  
\end{equation} 
where 
 \begin{equation}
  \label{TakibayevNZh_eq:44} 
  V^{ef}_{11} = \sum_{k,l\neq 1} G_1 |\nu_k > \eta_k \left( \eta^{-1}_k \delta_{kl} + P_{kl} \right) \eta_l < \nu_l |  G_0 t_1   \  .  
\end{equation} 
In the case when $V_1$ has a separable form, equations~(\ref{TakibayevNZh_eq:43}) and~(\ref{TakibayevNZh_eq:44}) can be reduced to~(\ref{TakibayevNZh_eq:10}) and~(\ref{TakibayevNZh_eq:11}).

Analytic solutions to the problem of light particle scattering on pair of interacting heavy particles are found in limit 
$\zeta\rightarrow 0$ and the method is stated in detail in  \cite{TakibayevNZh_PAN08}. 

It turns out that off-shell effects of light particle rescattering on heavy particles are contained only in the effective potential~(\ref{TakibayevNZh_eq:11}). That is, off-shell effects in the interaction of heavy particles are absorbed in the equation for the amplitude~(\ref{TakibayevNZh_eq:10}), where the effective potential is an on-shell quantity. 

\section{The scattering problem with two-body amplitudes of Breit-Wigner form}
\label{TakibayevNZh_sec:5}

 Now we consider the problem of neutron scattering on two fixed centers if the two-body scattering amplitudes have the  Breit-Wigner resonance form:
\begin{equation}
  \label{TakibayevNZh_eq:45} 
 t_i = - \frac{1}{\pi\rho (E)} \frac{\Gamma_2 /2}{E - E_{R,2} + i \Gamma_2 /4}   .  
\end{equation} 
The energy and width of this two-body resonance can be written through real and imaginary parts of resonance wave number: $E_{R,2} = (p_{R,2}^2 - p_{I,2}^2)/2m$ and 
$\Gamma_2 = 2 |p_{R,2} p_{I,2}|/m $. Here, index 2 marks two-body parameters (below index 3 will mark three-body parameters).  

It is known that the residue of resonant amplitude comes to factorization form in the region near  to point of polar singularity. In this region the representation~(\ref{TakibayevNZh_eq:37}) can be transformed to separable form~(\ref{TakibayevNZh_eq:2}). It is convenient to determine parameters of two-body  separable potential in terms of resonance energy and width of two-body scattering. 

We can normalize form-factors of two-body potential so that  
\begin{equation}
  \label{TakibayevNZh_eq:46} 
\nu(p) \approx \sqrt{\Gamma_2 /2\pi\rho(E)} \ ,   
\end{equation} 
if $p \approx  p_{R,2}$, and take hold of  
\begin{equation}
  \label{TakibayevNZh_eq:47} 
\eta^{-1}_i = - (E - E_{R,2} +i \Gamma_2/2)
\end{equation} 
instead of simple form~(\ref{TakibayevNZh_eq:3}).

So, exact solutions can be obtained in the case when pair t-matrices have  Breit-Wigner resonance form, too. For isolated pair resonance with energy $E_{R,2}$  and width $\Gamma_2$ we can get elements of $J$-matrix from~(\ref{TakibayevNZh_eq:13}) -~(\ref{TakibayevNZh_eq:15}) in analytic form.     

Therefore, zeros of  $D$-function  are determined by equation
\begin{equation}
  \label{TakibayevNZh_eq:48} 
 \left(E - E_{R,2} + i \frac{\Gamma_2}{2} - J \right)\cdot  \left(E - E_{R,2} + i \frac{\Gamma_2}{2} + J \right)= 0  .  
\end{equation} 
Then, introducing real and imaginary parts of $E = E_{res,3} = E_{R,3} - i \Gamma_3$ when $D = 0 $ and $J = J_R + i J_I$, we write out two branches of three-body resonances: 
\begin{equation}
  \label{TakibayevNZh_eq:49} 
 E_{R,3} = E_{R,2} - J_R , \qquad 
 \Gamma_3 = \Gamma_2 - 2 J_I , 
\end{equation} 
and 
\begin{equation}
  \label{TakibayevNZh_eq:50} 
 E_{R,3} = E_{R,2} + J_R  ,  \qquad
 \Gamma_3 = \Gamma_2 + 2 J_I  . 
\end{equation} 

Note that $J = J(b/2,p_0)$ has the oscillating behavior. For example, form-factors of two-body potentials in  ~(\ref{TakibayevNZh_eq:46}) acting in $S$-wave give 
\begin{equation}
  \label{TakibayevNZh_eq:51} 
 J_{ij} = J_{ji} = J = - \Gamma_2 \frac{\exp(ibp_0/2)}{bp_{R,2}} \  .  
\end{equation} 
Therefore, every branch contains sets of three-body resonances.    

Some of them are situated at energy scale above the energy of two-body resonance, and others - under this energy. Some of three-body resonances will have more narrowed width, others - more widen

Remarkably that there are several points where function $\Gamma_3 = \Gamma_3 (b/2,p_0) = 0$. It means that lifetime of resonance in these points becomes infinite. 
It is possible only in simple model, although in real situation lifetime of resonances can be enlarged substantially if distortions are suppressed.
  
Note, there are no principal difficulties to include more complicated forms of two-body separable potentials and other partial components into the model. 

At next subsections we consider the model of scatterer subsystem which consists of nuclei in the capacity of fixed centers. It is important that these nuclei have resonance interaction with neutron. And their two-body resonances give rise to three-body resonances. We investigate three-body resonance positions in dependence on $b$ - distance between centers. And we indicate specially points where imaginary part of resonance energy comes to zero.

\subsection{The scattering of neutron on two fixed $\alpha$-particles}
\label{TakibayevNZh_sec:51}

At first we consider the low energy resonance scattering of neutron on subsystem of two fixed $\alpha$-particles. 

It is known that two-body $n,\alpha$-system does not have bound or resonant states in $S$-wave at low energies as repulsive forces act between nucleon and $\alpha$-particle in this case. However, there are resonances in other partial waves.   

The respective amplitudes have resonances in the $P$-wave components $P^J_{L,S}$, where the total momentum $J = 3/2, 1/2$, orbital - $L = 1$, and $S = 1/2$. We take into account the distinct resonance $P^{3/2}_{1,1/2}$  which has parameters
$E_{R,2} \approx 0.9$ MeV  and $\Gamma_2 \approx 0.6 $ MeV \cite{TakibayevNZh_Ajz}.

This resonance can be described satisfactorily by the simple separable potential, for instance, 
involving the form factors (see~(\ref{TakibayevNZh_eq:5}))
\begin{equation}
  \label{TakibayevNZh_eq:52} 
 \nu_P (p) = Const\frac{p/\beta}{1 + p^2/\beta^2}  \  ,  
\end{equation} 
which give rise to  
 \begin{equation}
  \label{TakibayevNZh_eq:53} 
 \eta^{-1}_P = \lambda^{-1}_P + \frac{1-2k}{(1 - i k)^{-2}}  \  ,  
\end{equation} 
where $\eta_P = \eta^{3/2}_{1,1/2}(k)$, $k = p_0/\beta$, and $Const = \sqrt{8\pi/(2\mu \beta)}$. 

The two-body amplitude has the pole at the point 
$k = k_{res,2}$ , 
where $\eta^{-1}_P(k_{res,2}) = 0$.
 
The resonance parameters, $E_{R,2}$ and $\Gamma_2$, can be associated with potential parameters $\lambda_P$ and $\beta$. So, there are two relationships: $\lambda_P = -(1 + k_{I,2})$ and $k^2_{R,2} = - k_{I,2} (1 + k_{I,2})$, where  $k_{res,2} = \pm k_{R,2} + i k_{I,2}$. It follows that the values of $k_{I,2} = - 0.0256$, $k_{R,2} = 0.158$, from which we can determine parameters of the nuclear potential: $\lambda_P = - 0.974$ and $\beta = 1.175 fm^{-1}$.

Moreover, we can describe  $n,\alpha $-scattering in $S$-wave with simple separable potential~(\ref{TakibayevNZh_eq:28}), too. In this case (see ~(\ref{TakibayevNZh_eq:29})) the positive coupling constant gives the repulsive character of the $S$-wave interaction $\lambda_S = 15 $,
$ \beta_S = 1.4 fm^{-1}$ . 

Note that the enhancement factor in~(\ref{TakibayevNZh_eq:53}) can be presented in form similar to~(\ref{TakibayevNZh_eq:47})
 \begin{equation}
  \label{TakibayevNZh_eq:54} 
 \eta^{-1}_P = - (E - E_{res,2})\cdot A_P (E,E_{res,2})  \  ,  
\end{equation} 
where $E_{res,2} = E_{R,2} - i\Gamma_2 /2 $
 \begin{equation}
  \label{TakibayevNZh_eq:55} 
 A_P (E, E_{res,2}) = \int d\vec{p} 
  \frac{\nu_P (\vec{p}) \nu_P^* (\vec{p})}{(E - E_p + i0)(E_{res,2} - E_p)} \  \  ,  
\end{equation} 
and $E_p = p^2/2m $.

Thus, the matrix $J$ (see~(\ref{TakibayevNZh_eq:15})) contains submatrix  in respect of  indices of two-body partial waves - $S$ and $P$: \\
\centerline{ $J =$
\begin{math}
\bordermatrix{ & \cr 
  & 0 & J_{12}  \cr   
  & J_{21} & 0  \cr }
 \end{math}, \quad 
$J_{12} = J_{21} = J_{SP} = $  
\begin{math}
\bordermatrix{ & \cr 
  & J_{SS} & J_{SP}  \cr   
  & J_{PS} & J_{PP}  \cr }
 \end{math}.} \\
 
It is obvious that we can write $J_{SP}$-element as  
 \begin{equation}
  \label{TakibayevNZh_eq:56} 
 J_{SP}(p_0; \vec{r}) = \int d\vec{p} \exp(i \vec{r}\vec{p}) 
  \frac{\nu_S (\vec{p}) \nu_P (\vec{p})}{E - E_p + i0} \ , 
\end{equation} 
and other elements of $J_{SP}$-submatrix in analogous form. 

For $S$-wave components of $B$-matrix elements we have 
 \begin{equation}
  \label{TakibayevNZh_eq:57} 
B_{L,L'} = \sum_K J_{L,K} \ \eta_K (p_0) J_{K,L'} \ \eta_{L'}(p_0) \ , 
\end{equation} 
 where $L, L', K = S,P$. 
 
Then it is not difficult to determine $D$-function and calculate values of resonant energy $E_{res}$ in three-body system and $b$ - distance between two heavy centers. 
  
In framework of the method we obtained exact analytic solutions for "quasi-bound" states when the widths $ \Gamma_3 = 0 $. It is remarkable that near this points $p_{I,3} $ may cross zero and be even positive. 

Our calculations concerned a region of low energy only. Values of resonance energies - $E_{R,3}$ and parameters $b_k$ - distance between fixed $\alpha$-particles, when $\Gamma_3 = 0$, are given by Table 1.  
Note, that in the case of $\Gamma_3 = 0$ real and imaginary parts of $D$-function are both equal to zero. 

Simple calculations give two resonance states  with  $\Gamma_3 = 0$, named here as "quasi-bound" states. In Table~\ref{TakibayevNZh_tab:1}  resonance energies $E_{R,3;k}$ are given in MeV and  distances between two fixed $\alpha $-particles $b_k$ - in $fm$, where $k = 1,2$ - the number of resonance state.

\begin{table}[h]
\caption{Neutron and proton "Quasi-bound" states (with $\Gamma_3 = 0 $ ) 
in model of two fixed $\alpha $-particles.} 
\label{TakibayevNZh_tab:1}       
\begin{tabular}{lcccc}
\hline\noalign{\smallskip}
Three-body system  & \ \ \  $E_{R,3;1}$ \ \ \ & $b_1$ & \ \ \ $E_{R,3;2}$& \ \ $b_2$ \\
\noalign{\smallskip}\hline\noalign{\smallskip}
$n + \alpha, \alpha$ &   0.88 &  18  &   1.37 &  31 \\
$p + \alpha, \alpha$ &   1.78 &  14  &   2.36 &  24 \\
\noalign{\smallskip}\hline
\end{tabular}
\end{table}

Then we turn our attention to the proton scattering on subsystem of two fixed $\alpha$-particles. 
For comparison between parameters of ($n, \alpha, \alpha$) and ($p, \alpha, \alpha$) resonances the calculation data are shown in Table~\ref{TakibayevNZh_tab:1}.
Note that two-body $p, \alpha$ scattering amplitude has resonance in the $P$-wave too, with parameters: $E_{R,2} \approx 1.9$ MeV  and $\Gamma_2 \approx 1.5 $ MeV \cite{TakibayevNZh_Ajz}.  

The estimations of  ($p,\alpha,\alpha$) resonance parameters have been performed on the base of two-body resonance in $P$-wave  without ordinary repulsive Coulomb force between the proton and $\alpha$-particles in $S$ and $D$ waves. 

As a rule \cite{TakibayevNZh_BAZ,TakibayevNZh_PAN05} repulsive Coulomb forces result in widening of distance between centers and shifting of three-body resonance levels to higher energies. It is important that the difference of 1 MeV or more between resonance energies of   ($p,\alpha,\alpha$)-  and  ($n,\alpha,\alpha$)-systems is remaining.

\subsection{The neutron resonance scattering on subsystem of two fixed nuclei}
\label{TakibayevNZh_sec:52}

Now we consider the cases when more heavy nuclei than $\alpha$-particles are fixed in two-body scatterer subsystem. Subsystems with nuclei of atoms of oxygen and magnesium are interesting objects because they have similar features of interaction with neutron as $\alpha$-particle. 

The fact is that two-body scattering amplitudes of neutron on $^{16}O$ as well as on $^{24}Mg$ have
$P$-wave resonances at low energy region. Moreover, in these cases repulsive forces act in $S$-wave  like in $N, \alpha$-particle scattering amplitude.  

We take into account only the lowest isolated neutron-nucleus resonance states in two-body systems. The two-body resonance parameters  are $E_{R,2} = 435 $ keV, $\Gamma_2 = 40 $ keV in the case of ($n,  ^{16}O $), and $E_{R,2} = 84 $ keV, $\Gamma_2 = 13 $ keV in the case of ($n,  ^{24}Mg$). 

As above we describe  neutron-nucleus scattering in $S$-wave with simple separable potential~(\ref{TakibayevNZh_eq:28}) with the same positive coupling constant $\lambda_S = 15 $ and  $ \beta_S = 1.4 fm^{-1}$ . 

Two-body resonances in $P$-wave can be described with separable potentials  
involving the form factors~(\ref{TakibayevNZh_eq:52}), where potential and resonance parameters have to be coordinated with each other. It is resulted in $\lambda_P = 0.999$, $\beta_P = 6,076$ $fm^{-1}$ in case of $^{16}O$, and in $\lambda_P = 0.998$, $\beta_P = 1,635$ $fm^{-1}$ in case of $^{24}Mg$. 
 
Then we determine three-body quantities following the scheme of previous subsection.
The main aim is the determination of $D$-function and positions of zeros of this function. 

Calculations give the points where $\Gamma_3 = 0$. Energies of these quasi bound states of neutron in subsystem of two fixed nuclei and distances between these centers are shown in Table~\ref{TakibayevNZh_tab:2}. Note that $E_{R,3;k}$ are given in keV and $b_k$ - in $fm$, $k = 1,2$ - number of resonance state.   
 
\begin{table}[h]
\caption{Neutron "quasi-bound" states (with $\Gamma_3 = 0$) in model of two fixed nuclei. }
\label{TakibayevNZh_tab:2}       
\begin{tabular}{lcccc}
\hline\noalign{\smallskip}
Three-body system  & \ \ \  $E_{R,3;1}$ & \ \ \  $b_1$ & \ \ \ $E_{R,3;2}$ & \ \ \ $b_2$  \\
\noalign{\smallskip}\hline\noalign{\smallskip}
$n + (^{16}O , ^{16}O) $ &  231 &  10.53 &  286 &  12.18 \\
$n + (^{24}Mg , ^{24}Mg) $ &   82 &  63.27 &   117 &  70.37 \\
\noalign{\smallskip}\hline
\end{tabular}
\end{table}

It should be noted that nuclei with more heavy mass taken as fixed centers give rich and complicated pictures of three-body resonance states.

\section{The model of neutrino generator}
\label{TakibayevNZh_sec:6}

It is remarkable that the position of two-body $p,\alpha$ - resonance on energy scale is nearly 1 MeV above the position of $n,\alpha$ - resonance. Moreover, the resonances in ($p,\alpha,\alpha$)-system have the energies nearly 1 MeV above energies of corresponding resonances in ($n,\alpha,\alpha$)-system (see Table ~\ref{TakibayevNZh_tab:1}). 

Thus, if certain systems exist or can be made to keep these resonance states together the observation of transitions between these states becomes possible. 

It seems that this kind of systems can be formed during the evolution of cold stars, for example, in white dwarfs or other superdense stars. 
Indeed, the core of these stars can transform into solid crystalline body under very high pressure \cite{TakibayevNZh_Kirzh}.  
  
In this connection we consider the model of ideal crystal where nuclei are fixed in nodes of the lattice. It is supposed that distances between nodes of this perfect crystalline lattice can be changed and become small, for instance, as a result of very big pressure from outside or force of gravity \cite{TakibayevNZh_VolKir}. 

Now we consider the crystal model of stellar Helium core. This simple model may be interesting in astrophysics because it can play the role of low energy neutrino generator. The fact is that in this model the three-body resonance energies and widths are functions not only of two-body interaction parameters but also of lattice parameter $b$ - distance between nodes.  

It is very important that there are some values of parameter $b_k$ for which certain widths of three-body resonance states are close to zero (see, Table~\ref{TakibayevNZh_tab:1}).

For example, a system consisting of one neutron and two fixed $\alpha$-particles has the resonance energy  $E_{R,3} \approx 1.37$ MeV with  $\Gamma_3 \approx 0$ when  $b \approx 31$ $fm$. The analogous  system consisting of one proton and two fixed $\alpha$-particles has the resonance too, with the energy $E_R \approx 2.4 $ MeV and small width near this $b$.  

We suppose that distance between nodes of $\alpha$-particle crystalline lattice $b$ is close to $b_1 = 31$ $fm$. 

Here one might ask: How can neutron appear in the lattice?  

Note that the part of protons can penetrate inside crystal from external surroundings if they have enough energy for channeling in the lattice. Furthermore,  some  protons can be inside of lattice ever since the time of lattice formation \cite{TakibayevNZh_Kirzh}.

Then, owing to reaction $p + e^- \rightarrow n + \nu$  protons can turn into neutrons because it is the exoenergic reaction. So, neutrons appear in the $(n,\alpha,\alpha)^*$- quasi-bound states.  

Neutrinos with energy $E_{\nu} \approx 0.2$ MeV are produced in this reaction, after what they leave the lattice. 

This $(n,\alpha,\alpha)^*$-quasi-bound state has $\Gamma_3 \approx 0$ and very big lifetime if  $b \approx 31$ $fm$. Besides, the decay of neutron is suppressed in the crystal. 

In the case of distortions the condition in the lattice for existence of this neutron  states is not supported and via $\beta$-decay the neutron turns into proton, producing electron and antineutrino. 

So, we assume that the following reaction can be stimulated in the crystalline lattice:
 \begin{equation}
  \label{TakibayevNZh_eq:58} 
 (p,\alpha,\alpha)^*_{2.36 MeV} + e^- \rightarrow (n,\alpha,\alpha)^*_{1.37 MeV} + \nu_e  \ ,  
 \end{equation} 
 and then
 \begin{equation}
\label{TakibayevNZh_eq:59} 
 (n,\alpha,\alpha)^*_{1.37 MeV} \rightarrow n + (\alpha,\alpha) \rightarrow p' + e^- + \bar{\nu}_e + (\alpha,\alpha) \nonumber\\ .
\end{equation} 
Here, ($^\ast$) marks the quasi-bound state of nucleon in ($N,\alpha,\alpha$) system, and its resonance energy is shown in subscript.
Note, that $E_{p'} \approx E_p - 2.3$ MeV.
   
The lattice distortions may be created periodically by satellites of star.  
As a result the star will generate clouds of neutrinos spreading outside in space.  
In the case when neutrons can reach the star atmoshere they stimulate nuclear reactions.
   
The situation with neutrinos generation can be similar in the cases of stellar  crystalline cores of  $ ^{16}O$ and especially  $ ^{24}Mg$,  and more heavy nuclei.

\section{Conclusion}
\label{TakibayevNZh_sec:7}

The exact analytical solutions have been considered above in three-body problems, when one light particle interacts with subsystem of two heavy particles fixed in coordinate space. It is important that solutions of the problem can be followed out, i.e. obtained in analytic forms. 

These solutions can show the main three-body characteristics, for example, amplitudes, resonance states and their dependences on $b$ - distance between scatterer centers. Thus, exact solutions can be used as principle approximations in many problems of three-body quantum mechanics. 

It would be remarkable to create a setting, where $b$ - distance between nodes, could be changed in order to investigate properties of neutron resonance scattering on different monocrystals. Of course, this setting should operate with special monocrystals kept under very high pressures and very low temperatures. 

In this setting transitions between three-body $p$ and $n$ resonance states in monocrystal could be determined. It means to discover new neutrino generator and possibly the new mode of star cooling without nuclear reactions.

The analogous model of exotic particle interactions with the nodes of quark lattice or substructure may be interesting. In this case huge mass of heavy neutrinos or exotic neutral particles generated by star or galaxy center would be expanding and increasing far out to the galaxy frontiers and cripple the motion of satellites. Similar models might be useful for solving the problem of dark matter.

In any case it is clear that three-body effects and three-body resonances as well as quantum mechanics of three-body systems on the whole will give an important contribution in modern astrophysics.

\end{document}